\documentclass[preprint, 12pt]{elsarticle}

\DeclareGraphicsExtensions{.png,.jpg}
\usepackage{graphicx}

\usepackage{amssymb}

\usepackage{algorithmic}
\usepackage{algorithm}

\def\be{\begin{equation}}
\def\ee{\end{equation}}
\def\ba{\begin{array}}
\def\ea{\end{array}}
\def\bea{\begin{eqnarray}}
\def\eea{\end{eqnarray}}
\def\beas{\begin{eqnarray*}}
\def\eeas{\end{eqnarray*}}

\newcommand{\bfx}{\mbox{\boldmath $x$}}

\newcommand{\bfA}{\mbox{\boldmath $A$}}

\newcommand{\bfJ}{\mbox{\boldmath $J$}}

\newcommand{\bfT}{\mbox{\boldmath $T$}}

\newcommand{\bfxi}{\mbox{\boldmath $\xi$}}

\journal{Computers and Mathematics with Applications}

\begin{document}

\begin{frontmatter}



\title{Vectorized OpenCL implementation  of  numerical integration for higher order finite elements}


\author[label1]{Filip Kru\.{z}el}
\ead{fkruzel@pk.edu.pl}
\author[label2]{Krzysztof Bana\'{s}\corref{cor1}}
\ead{pobanas@cyf-kr.edu.pl}

\address[label1]{
Institute of Computer Science, \\
Cracow University of Technology, Warszawska 24, 31-155 Krak\'{o}w, Poland }
\address[label2]{
Department of Applied Computer Science and Modelling, \\
AGH University of Science and Technology, Mickiewicza 30, 
30-059 Krak\'{o}w, Poland}

\cortext[cor1]{Corresponding author}

\begin{abstract}
In our work we analyze computational aspects of the problem of numerical integration in finite element calculations and consider an OpenCL implementation of related algorithms for processors with wide vector registers. 

As a platform for testing the implementation we choose the PowerXCell processor, being
an example of the Cell Broadband Engine (CellBE) architecture. Although the processor is considered old for today's standards (its design dates back to year 2001), we investigate its performance due to two features that it shares with recent Xeon Phi family of coprocessors: wide vector units and relatively slow connection of computing cores with main global memory. The performed analysis of parallelization options can also be used for designing numerical integration algorithms for other processors with vector registers, such as contemporary x86 microprocessors.


We consider higher order finite element approximations and implement the standard algorithm of numerical integration for prismatic elements. Original contributions of the paper include the analysis of data movement and vector operations performed during code execution. 
Several versions of the implementation are developed and tested in practice.

\end{abstract}


\end{frontmatter}


\section{Introduction}


\subsection{Trends in microprocessor design}
The development trends in microprocessor architecture show two different ways of scaling. The first is to increase the number of cores. For massively multi-core microprocessors, such as GPUs, the cores are relatively simple, offering lower performance than standard general purpose cores, while the resources per single core, e.g. in the form of fast memories, are usually small. Access times associated with variables in executed programs depend on the storage locations for variables, with three levels of memory usually explicitly available to programmers: registers, fast memory (shared by several threads) and slower global memory \cite{cuda_guide,AMD_APP_guide}.

Another trend is to increase the width of vector execution units within single processor cores. This trend is visible e.g. in the recent general purpose x86 cores (256-bit vector units in Intel Sandy Bridge and Haswell cores) and in the Xeon Phi co-processor \cite{Xeon_Phi_book}, having cores equipped with 512-bit wide vector units. In co-processors, the cores are less numerous than in GPUs, dozens as opposed to hundreds, and their complexity lies in between standard, general purpose cores (with out-of-order execution, branch prediction etc.) and simple GPU cores, with e.g. no instruction decoding units.

The PowerXCell processor that we consider specifically in the current paper can be considered as the predecessor of architectures with wide vector units. It has 128 vector registers, having width of 128 bits, and no scalar registers. Its performance depends heavily on the proper vectorization of the code. 

The CellBE architecture  \cite{CBE_Handbook} is well known in HPC community mainly due to being the essential ingredient of the first petaflops system, the Roadrunner computer developed by IBM in 2008. The CellBE architecture  \cite{CBE_Handbook} has attracted the interest of many researchers, who investigated its performance in several application domains. In the context of finite element calculations, the most important research concerned linear equations solvers and their computational kernels \cite{cell_kernels,RojekS09,Kushida_FEM_Cell,Bepop_1}.

Both contemporary microprocessor architectures presented above, GPUs and Xeon Phi, share the same co-processor design, with slow PCI Express bus connecting coprocessors with host CPU and its memory. Similarly, the throughput of PowerXCell connection to DRAM memory is much lower than the combined throughput of links between PowerXCell vector cores and their local memories.  For such architectures, data movement between different levels of memory may become an issue of primary importance \cite{Govindaraju_06}. Access times to different levels of memory can differ by several orders of magnitude, so an improper design of algorithms can result in slow execution times.



In consequence, many existing algorithms have to be redesigned for new architectures. Redesigning should begin with a proper analysis of an algorithm, indicating operation count and detailed data movement requirements. When counting operations one should take into account whether operations are performed by scalar or vector units, while the analysis of data movement should incorporate different levels of memory hierarchy. The designs can be different for different architectures, since optimization performed for one architecture may turn out to be ineffective for another \cite{Rul_10}.


\subsection{Finite elements on modern computer architectures}
Numerical integration is the part of FEM codes where local, element stiffness matrices are calculated for a given weak statement of the problem solved. The matrices are then assembled into a global system of linear equations which is solved subsequently.

The time required for the solution of the system of linear equations usually strongly dominates the time of the whole FEM solution process. This justifies the fact that most of research in the area of porting finite element codes to new processor architectures concentrates on solvers of linear equations \cite{goddeke_solid,GoStMo_07scalability}. Often the main stress is put on computational kernels of linear solvers that are properly optimized for particular architectures \cite{Volkov_2008,cell_kernels,RojekS09}.

However, the time necessary for numerical integration should not be neglected. Computational complexity estimates, that will be presented later in the paper, show that in some cases the time required for numerical integration can be comparable to the time for solving the system of linear equations (or even larger for iterative solvers). Linear equations solvers employed in FEM codes are often highly optimized for each appearing computer architecture. If such a module is combined with an improperly designed and implemented numerical integration algorithm, the latter can become a serious performance bottleneck.

The relative significance of the performance of numerical integration depends on the problem (its weak statement) and the approximation applied. For low order finite element approximations, e.g. linear elements, the time required for element stiffness matrix creation is usually short, especially when compared to the time necessary to solve the associated system of linear equations. For many simple problems, integrals can even be precomputed analytically and the creation of stiffness matrix entries changes into purely algebraic operations, with closed formulas into which element parameters, like e.g. vertices coordinates, are substituted. In such cases 
creating the
global stiffness matrix consists mainly of assembling, performed for quickly calculated element stiffness matrices \cite{goddeke_solid}.

The situation is different for higher order approximations where polynomials with high degree $p$ are used. 
In such cases special element shape functions are often applied and special integration methods designed \cite{dubiner,sherwin}. 
These special methods can be relatively simple, as e.g. for hexahedral elements with tensor product shape functions, where standard $O(p^9)$ complexity of numerical integration can be reduced to $O(p^7)$ \cite{leszek_hp_book_v2}. For other types of elements, it is also possible to achieve lower complexity, but the costs associated with integrating special shape functions make the algorithms less efficient than standard techniques for lower orders of approximation \cite{schwab_integr,Vos_2010}.

Since low order polynomials are the most popular in practice, investigations concerning the creation of finite element stiffness matrices on modern processor architectures often concentrate on assembly procedures \cite{Markall_2010,Cecka_2011,Cecka_gems,Markall_2013,Knepley_2013}. The cited articles all consider assembly on GPUs, the authors of the current paper are not aware of any attempts to explicitly vectorize assembly procedures on PowerXCell or other processors with wide SIMD registers. 

For higher order approximations, investigations concerning the porting of finite element codes to new computer architectures also focus mainly on GPU implementations \cite{komatitsch09,Klockner_2009,Filipovic_assembly,Dziekonski_assembly,Dziekonski_generation}. Although several issues are common to PowerXCell and GPU implementations, the general design philosophy and flow of execution are different.

One should also mention here attempts to create formal specifications of finite element calculations, from which procedures for particular architectures are automatically created by suitable compilers \cite{Kirby_compiler,Logg_DOLFIN,Fenics_book,Markall_2010,Markall_2013,vandeGeijn_1,vandeGeijn_2}. Although very interesting in themselves, the specifications usually embody substantial knowledge concerning problems, algorithms and hardware in question. They can be considered as a further step in the investigations on the interplay between algorithms and computer architectures, that we try to develop in the current paper.

\subsection{Current contribution}
In the present article we consider a PowerXCell implementation of the standard finite element numerical integration technique for 3D problems and orders of approximation up to 7. 
The limiting order 7 is chosen, to certain extent, arbitrarily. For low order approximations (usually up to 3) the standard integration technique is more efficient and simpler than special techniques \cite{Vos_2010}. 
For higher orders, it can be, in some cases, profitable to switch to special techniques of integration even for orders in the range 4-5 \cite{schwab_integr,Vos_2010}. The optimal integration technique may be different for different problems (weak statements) and different approximation methods.

We choose standard integration techniques with an eye to applicability to different types of elements and the generic character of the approach, valid for different finite element formulations and orders of approximation.
We do not consider particular weak statements for different problems in science and engineering, instead, we concentrate on a single term that appears in many problems.

The present article is devoted to a thorough analysis of computational aspects of standard finite element numerical integration and the implementation of related algorithms for the CellBE architecture, in particular the PowerXCell processor. It is a continuation and extension of works presented in \cite{ppam_09_cell}. 

As new developments, we present the analysis of data movement during calculations, that leads to the selection of implementation strategy. Moreover, we show how vector capabilities of current microprocessors can be utilized for numerical integration calculations.

We present the results of computational tests that illustrate the performance of created implementations. The results
were obtained using a modular finite element code ModFEM, developed for standard and discontinuous Galerkin finite element simulations \cite{iccs04, iccs_10, cmms_modfem_2013}. The entire work reported in the paper is undertaken in an effort to port the code to modern computer architectures, including GPUs and many-core processors.

We are not aware of any other investigations concerning finite element numerical integration on the PowerXCell processor and, in a broader context, the explicit use of vector capabilities of modern processors in finite element numerical integration procedures.


\section{Finite element numerical integration}
\label{section_num_int}
Calculation of integrals from a weak statement in finite element codes
is usually performed assuming that the whole computational domain, as the domain of
integration, is divided into finite elements, of one or several types
--- triangles, quadrilaterals, tetrahedra, prisms, hexahedra, etc.
For distributed memory implementations, there exists also a partition of the computational domain into subdomains, with each subdomain associated with a single process performing calculations. When considering such domain decomposition approach, numerical integration appears as an embarrassingly parallel algorithm (there are no dependencies and for proper overlapping decompositions there is also no communication necessary) \cite{ppam03,dd15}. In the rest of the paper we assume the domain decomposition technique for distributed memory machines and for further parallelization we concentrate on a single subdomain and a single CPU process.

For a single subdomain, in a loop over all subdomain elements, at each element, integrals corresponding to pairs of finite element basis functions are computed and the results stored in local, element stiffness matrices. Local load vectors are obtained through integration of corresponding terms as well, but this procedure is computationally much less demanding and we neglect it in the present article. 


Computed stiffness matrices are then assembled into a global stiffness matrix, the matrix of the associated system of linear equations. A single entry in the global matrix can be a sum of entries from several local matrices. Hence, the assembly cannot be considered as an ''embarrassingly'' parallel algorithm with no dependencies.

However, classical techniques of coloring can be used to remove dependencies. The crucial observation is that two local matrices contribute to a single global entry if there is a neighborhood relation between them \cite{leszek_hp_book_v1}. Using a proper coloring scheme, one can obtain sets of elements with different colors, with two elements having the same color if their local stiffness matrices do not coincide at any global matrix entry. Given this partition of elements, it is assumed that the algorithm of numerical integration (followed by immediate matrix assembly) proceeds color by color. Since we consider large scale calculations, it can be safely assumed that each set of elements of the same color has sufficient number of elements, in order to fully utilize parallel capabilities of the hardware. Hence, when considering creation of local stiffness matrices by a single microprocessor, one can assume that there are no dependencies during assembly procedure. In consequence, numerical integration for different elements can also be considered as dependence free and perfectly parallelizable.

We consider numerical integration for an example second order term with partial derivatives of trial and test functions (the term that appears e.g.~for approximations of Laplace operator). The formula for obtaining an entry $A^e_{ij}$ to the local stiffness matrix $\bfA^e$, associated with element $\Omega_e$ can be expressed as:
\be
A^e_{ij} = \int_{\Omega_e}\sum_{k=1}^3
\frac{\partial\phi_i}{\partial x_k}
\frac{\partial\phi_j}{\partial x_k}
d\Omega
\label{wzor1}
\ee
where $\phi_i$ and $\phi_j$ are global basis functions. 

In order to compute integrals of the form (\ref{wzor1}) the change of variables is applied, which in practice means that we always perform integration on a reference element of a particular type. The transformation from the reference element to the real element is denoted by $\bfx(\bfxi)$. For the reference element we use its shape functions 
$\hat {\phi}_i$, instead of global basis functions $\phi_i$, and apply some form of numerical quadrature. The numerical quadrature transforms the integral into a sum over integration points within the reference domain. 

From different possible quadratures we concentrate on the most popular Gaussian quadratures. 
We assume that we use ${N_{\rm{I}}}$ integration points with local coordinates $\bfxi^I$ and the associated weights $w^I$. The number of integration points is determined by the requirement to accurately compute products of shape functions, neglecting non-linearity of coefficients and element geometry. This should suffice for the convergence of finite element approximations \cite{ciarlet}. After performing the steps described above, integral (\ref{wzor1}) is transformed to the sum:
\be
A^e_{ij} \approx \sum_{I=1}^{N_{\rm{I}}} 
\sum_{k=1}^3
\frac{\partial \hat {\phi}_i}{\partial x_k}
\frac{\partial \hat {\phi}_j}{\partial x_k}
\det \bfJ_{\bfT_e}w^I
\label{wzor2}
\ee
where  $\bfJ_{\bfT_e}$ is the Jacobian matrix of transformation $\bfx(\bfxi)$. 

The number of shape functions for an element, $N_{\rm{sh}}$ (the range of indices $i$ and $j$ in (\ref{wzor1}) and (\ref{wzor2})) depends on the degree of approximating polynomials $p$. For several typical 3D elements it is given in Table \ref{tablica_1}. In the table, complete basis and tensor product basis refer to the way shape functions are defined for an element \cite{solin} (for prismatic elements shape functions are obtained as products of complete basis for horizontal triangular faces and 1D shape functions along the vertical direction). In any case it can be seen that the number of shape functions is $O(p^2)$ for 2D elements and $O(p^3)$ for 3D elements. According to our assumptions specified above, the number of Gauss points is of the same order.
\label{section_accuracy}

\begin{table}[t!]
\begin{center}
\begin{tabular}{|c|c|}
\hline
Type of element & Number of shape functions \\
\hline
Triangular (complete basis) &  $\frac{1}{2}(p+1)(p+2)$  \\
\hline
Quadrilateral (tensor product basis) &  $(p+1)^2$   \\
\hline
Tetrahedral (complete basis) & $\frac{1}{6}(p+1)(p+2)(p+3)$ \\
\hline
Prismatic  & $\frac{1}{2}(p+1)^2(p+2)$ \\
\hline
Hexahedral (tensor product basis) & $(p+1)^3$ \\
\hline
\end{tabular}
\end{center}
\caption{The  number of shape functions 
  associated with a single element of a given type for different orders of approximation $p$.} 
\label{tablica_1}
\end{table}

\begin{algorithm}
\caption{Standard sequential finite element numerical integration algorithm.}
\label{alg_seq}
\begin{algorithmic}[1]
\FOR{$ielem=1$ \TO $nr\_elems\_in\_subset$}
\STATE read geometry data for element
\STATE prepare quadrature data, $\bfxi$ and $w$, for all integration points
\STATE initialize element stiffness matrix, $\bfA^e$
\FOR{$I=1$ \TO $N_{\rm{I}}$} 
\STATE read or compute values of shape functions and their derivatives with respect to local element coordinates
\STATE read or calculate Jacobian matrix, its determinant ($\det \bfJ_{\bfT_e}(\bfxi^I)$) and its inverse
\STATE calculate derivatives of element shape functions with respect to physical coordinates $\bfx$, $\frac{\partial \hat {\phi}_j}{\partial x_k}(\bfxi^I)$
\FOR{$i=1$ \TO $N_{\mathrm{sh}}$}
\FOR{$j=1$ \TO $N_{\mathrm{sh}}$}
\FOR{$k=1$ \TO 3}
\STATE $A^e_{ij} +\!\!=  \frac{\partial \hat {\phi}_i}{\partial x_k}(\bfxi^I)  \times \frac{\partial \hat {\phi}_j}{\partial x_k}(\bfxi^I)  \times 
\det \bfJ_{\bfT_e}(\bfxi^I) \times w^I $
\ENDFOR
\ENDFOR
\ENDFOR
\ENDFOR
\ENDFOR
\end{algorithmic}
\end{algorithm}

\subsection{Algorithm of numerical integration}
Algorithm~\ref{alg_seq} represents a sequential numerical integration scheme for creating local, element stiffness matrices, corresponding to formula (\ref{wzor2}). We assume that calculations are performed in several loops with the outermost loop being the loop over elements of a single color (if coloring is used) and over a single subdomain (or the whole computational domain).

In the algorithm, a popular ordering of calculations is assumed, with the loop over integration points moved outside the loops over degrees of freedom (shape functions). At each integration point, the values of all shape functions and their derivatives are computed or read from precomputed tables. Thanks to this, the values reside in some faster memory and are efficiently utilized in calculations of stiffness matrix entries.

\subsection{Computational and memory complexity}
We perform the analysis of computational and memory complexity of the presented algorithm of numerical integration considering additionally necessary data movements between memory and processors/cores performing calculations. We assume that, at the beginning, the data is stored in some global memory. This data include information on elements, weak formulation and numerical quadratures. 

As we already pointed out, from the computational point of view, numerical integration for different elements is performed independently. Hence, the number of operations and the number of memory transfers grow linearly with the number of elements, ${N_{\rm{e}}}$. We concentrate next on the analysis for a single element. The analysis concerns subsequent steps (identified by the line number) from Algorithm~\ref{alg_seq}.

In step 2, data concerning element geometry is transferred from global memory to some faster memory available to threads. The size of read data structures depends on the type of element (linear, multi-linear or curvilinear, tetrahedral, prismatic, hexahedral, etc.). In our estimates, we assume that this size is a constant parameter (i.e. we do not consider isoparametric elements, with the number of geometry parameters growing with $p$). 
  
In the next step, the values of quadrature data are read from precomputed tables stored in global memory. The number of quadrature points depends on the weak statement (the degree of non-linearity of the integrated terms) and the order of approximation. We assume, as it was already stated, that the number of quadrature points is of order $p^2$ for 2D elements and $p^3$ for 3D elements. For each
reference element of a given type and given degree of approximation quadrature data are the same. Hence it is advantageous to put these data in some fast memory (which usually is small). If all elements have the same type and degree of approximation the data should be kept there during the whole procedure (i.e. for the whole loop over elements for which entries to the global stiffness matrix are calculated). Otherwise, one can organize the loop over elements in such a way, that elements are grouped into sets of elements with the same type and degree of approximation $p$ and in this way, minimize the required number of retrievals of integration points data from some large slow memory.

Initialization of the local stiffness matrix performed in step 4, with some negligible cost as compared to proper calculations, concludes the preparatory phase of single element calculations.

The three steps that we consider next, steps 6, 7 and 8, are performed inside the loop over integration points. First, the values of all shape functions and their derivatives with respect to local coordinates are either computed or retrieved from precomputed tables. 
The number of computed or retrieved values can be easily calculated for different types of elements and different orders of approximation $p$ based on data from Table \ref{tablica_1}. 

In step 6 of Algorithm ~\ref{alg_seq} only the values at single integration point are necessary. However, these values are the same for all elements of a given type and degree of approximation.
If the fast memory available to processors/cores is large enough, than the values for all shape functions and all integration points can be stored there and reused for subsequent elements. If the fast memory is too small for that, the values have to be retrieved from the global memory (or recomputed) for each integration point.

The cost of computing the values for a single shape function and their derivatives depends on the type of basis for the space of polynomials and a particular way of performing calculations, where several techniques for reusing values can be utilized. This cost has to be compared with the cost of retrieving the data either from the fast, local memory or the global memory (if the fast memory is too small).
The cost of calculations for each shape function grows with the growing polynomial degree, hence the option of precomputing becomes more attractive for higher orders of approximations. Additionally, one can take into account the fact that the importance of costs associated with shape functions evaluations within the whole numerical integration algorithm diminishes with the increasing degree $p$, due to the growing number of iterations over shape functions in steps 9 and 10.

At each integration point also the Jacobian matrix of transformation $\bfx(\bfxi)$ (and its inverse, necessary to compute derivatives of shape functions with respect to physical coordinates) is calculated in step 7 of Algorithm ~\ref{alg_seq}. The complexity of these operations depend on the way the geometry of the element is described. For simple geometries, that we assume in our analysis, the cost is constant (in the range of tens of operations). 
 
Finally, in a double loop over shape functions, all the entries of the element stiffness matrix are updated. There are $N_{\rm{sh}} \times N_{\rm{sh}}$ iterations and in each iteration one entry is updated with suitable values corresponding to a particular integration point. We assume a constant cost of updating a single entry. In our example term from formula \ref{wzor2} there are exactly 7 operations per single entry (after some obvious optimizations of the formula in step 12 of  Algorithm ~\ref{alg_seq}).

The total number of iterations in the triple loop (over integration points and shape functions) of numerical integration algorithm is $O(p^6)$ for 2D elements and $O(p^9)$ for 3D elements. This is the reason why for high orders of approximation, execution time can be comparable or even longer for numerical integration than for solving the associated system of linear equations. In the 3D case, the cost of solving the system of equations for sophisticated direct solvers is $O(N_{\rm{e}} p^9+N_{\rm{e}}^2p^6)$ \cite{maciek_solver_1, maciek_complexity_1} and for iterative solvers (if they converge) is usually lower.

In the following sections we show the design of an implementation of numerical integration for the PowerXCell processor, taking into account the analysis performed above. The design of implementations for graphics processors is presented in the companion article \cite{cmwa_12_ni_GPU}.  

\section{PowerXCell 8i}
\begin{figure}[t!]
\centering
\includegraphics[width=12cm]{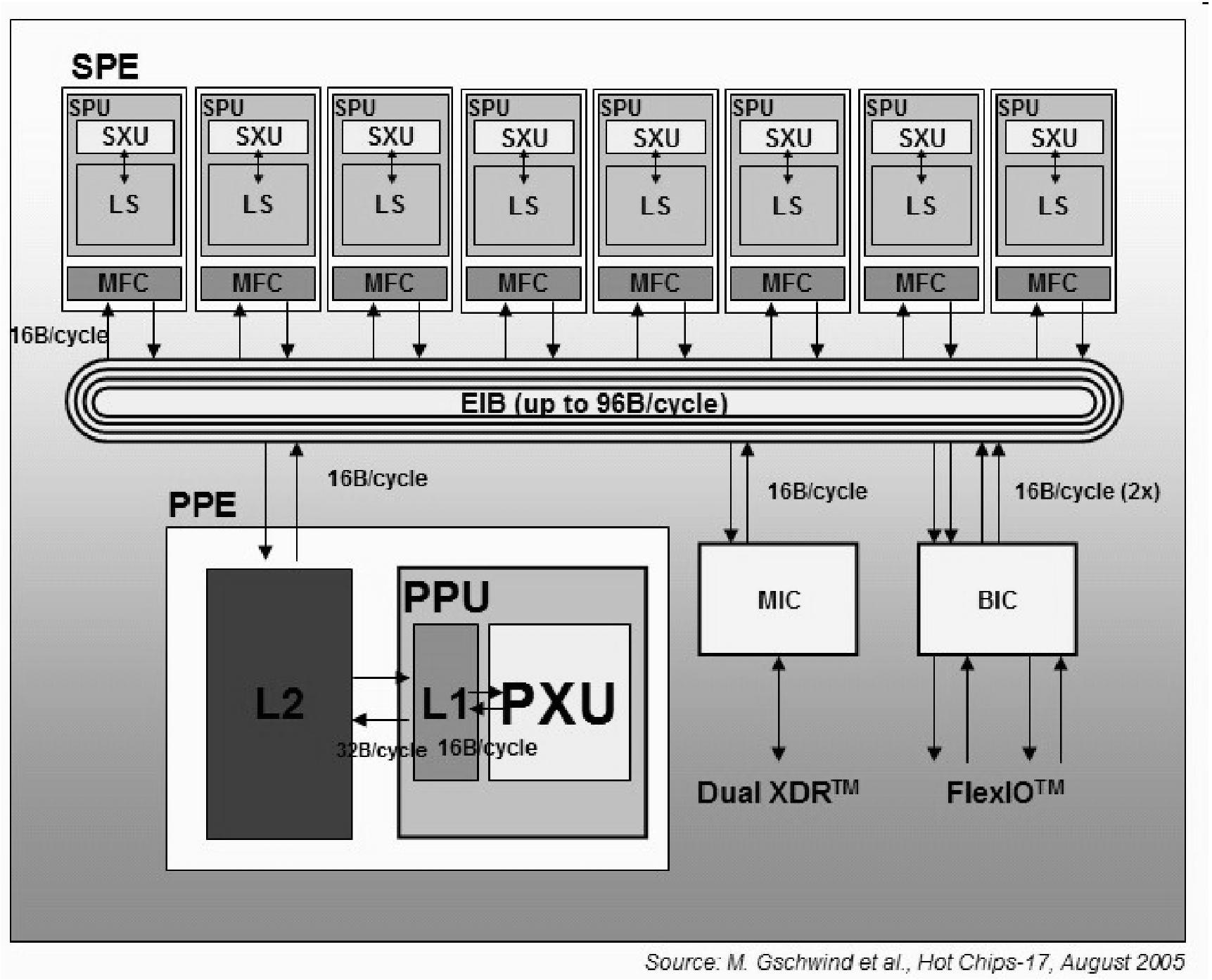}
\caption{CellBE architecture.}
\label{obrazek4}
\end{figure}

As a hardware platform for computations we use IBM PowerXCell 8i processor (Fig.~\ref{obrazek4}), which is a variation of the CellBE architecture \cite{CBE_Handbook} enhanced with double precision computations. This architecture combines  one standard general purpose IBM PowerPC core (called PPE -- Power Processor Element) with eight small, vector SIMD cores (called SPE -- Synergistic Processor Elements). Different parts of the Cell architecture are responsible for different tasks -- PPE performs operating system tasks and manages the execution of standard and SPE programs, while SPEs efficiently execute data processing algorithms. 


Figure ~\ref{obrazek4} presents a schematic diagram of CellBE architecture with indicated data transfer rates between different elements. These rates give the bandwidths of 25.6 GB/s for accessing L2 and global memories. 

Each SPE core is equipped with 128 128-bit registers, 256KB of fast private memory (local store, LS) and has a SIMD instruction set. It has one execution unit with two pipelines, one for integer and floating point operations and the second for e.g. load and store instructions. It can issue one instruction for each pipeline at every cycle. In particular it can issue one FMA (fused multiply-add) instruction for SIMD registers. Since one FMA corresponds to 8 standard single precision arithmetic operations, this leads to the theoretical maximum performance of 25.6 GFlops (single precision) for the 3.2 GHz clock. 

At each cycle a SPE core can load or store 16 bytes (the content of one 128-bit register) from/to its local memory. Careful implementation of algorithms for which the ratio of local memory references to floating point operations is lower than one and these two operations use most of the execution time, should lead to performance levels close to the theoretical maximum (this has been shown e.g. for matrix-matrix multiplication routines \cite{rw_mat_mul}).

\subsection{Programming model for PowerXCell}
Programming model for PowerXCell architecture can be chosen as either defined by native CellBE Software Development Kit (CellBE SDK) \cite{CBE_Handbook} or by OpenCL specification \cite{OpenCL}. 
In the current paper we consider OpenCL implementation only. For our initial experiments with the native Cell/BE SDK we refer to \cite{ppam_09_cell}.


The platform model of OpenCL was developed mainly for GPUs and consists of a host which is connected to one or more OpenCL devices. The devices are composed of compute units, which contain processing elements performing the proper calculations. For PowerXCell, the model is simplified: the role of the host is played by PPE, while each SPE can be identified with a compute unit having a single processing element. All SPEs form one device, with no global memory.

Similarly, the complex organization of execution in OpenCL, where threads (called work-items in OpenCL nomenclature) are grouped into work-groups, is simplified for PowerXCell. There are work-groups assigned to different SPEs, but in a fashion natural for standard cores, each work-group is composed of a single thread. Threads execute SPE programs, called kernels, written in OpenCL language being a variation of C.

This rich memory hierarchy of OpenCL (host, global, constant, local and private memories \cite{OpenCL}) is simplified when designing implementations for PowerXCell. There is no direct counterpart of constant memory, the host memory and the global memory of the device are identified with PowerXCell DRAM. 
The variables explicitly designed for OpenCL local memory are stored in
SPEs local memories. Contrary to the native Cell programming model, in OpenCL it is possible to write kernels without explicit memory transfer operations. The compiler translates operations of writing to and reading from global memory into suitable low level transfers over EIB bus.

There are two important ingredients of OpenCL programming model that are vital for programming PowerXCell and that are absent in standard programming languages. The first is the explicit management of memory hierarchy and the second is the explicit use of vector variables and vector registers.



\section{Implementation of numerical integration algorithm for the PowerXCell processor}
There are several levels of parallelism available for execution of programs on contemporary computer systems. We consider the mapping of numerical integration algorithm to hardware, taking into account all levels explicitly under the programmer's control.

The first, outermost level is the level of nodes of distributed memory parallel machine. As we already stated, this level is fully exploited by the domain decomposition approach. 

The next level of parallel hardware organization is the level of processors/cores. For OpenCL and PowerXCell this again is simple, with single threads running on SPEs, that play the dual role of compute units and processing elements. Again, we design the parallel version of the algorithm assuming the mapping of iterations in the loop over elements to SPEs, with a single SPE thread performing integration for a subset of elements. 

The only difference between OpenCL programming for SPEs and standard (e.g. OpenMP for x86 cores) programming,  notable at this level, is the small size of available fast memory for threads. For standard programming, the existence of several levels of cache makes the problem of data localization less significant than for SPEs, equipped with non-cached local stores. For PowerXCell the use of local memories must be explicitly designed, taking into account the size of data structures and the cost of memory transfers.

For the purpose of detailed derivation of parallel implementation, we select a particular type of element: the standard geometrically bilinear prismatic element. This element is the simplest one requiring the evaluation of Jacobian at each integration point, assumed in Algorithm \ref{alg_seq}.

\begin{table}[t!]
\begin{center}
\begin{tabular}{|l|r|r|r|r|r|r|r|}
\hline
Stored &  \multicolumn{7}{|c|}{Degree of approximation p}  \\
quantities & 1 & 2 & 3 & 4 & 5 & 6 & 7   \\
\hline
$\hat{\phi}_i$, $\frac{\partial \hat{\phi}_i}{\partial \bfx}$
&24&	72&	160&	300&	504&	784&	1152\\
\hline
$\bfxi^I$, $w^I$
&24&	72&	192&	320&	600&	924&	1344\\
\hline
$\bfA^e$
&36&	324&	1600&	5625&	15876&	38416&	82944\\
\hline
$\hat{\phi}_i$, $\frac{\partial \hat{\phi}_i}{\partial \bfx}$ at all $\bfxi^I$
&144&	1296&	7680&	24000&	75600&	181104&	387072\\
\hline
\end{tabular}
\end{center}
\caption{The size of data structures -- in the number of single or double
  precision scalars -- for storing values used in numerical
  integration for 3D prismatic elements with different
  orders of approximation.}
\label{tablica_3}
\end{table}

Table \ref{tablica_3} presents the sizes of main data structures appearing in Algorithm \ref{alg_seq} when using prismatic elements. The entries are computed assuming:
\begin{itemize}
\item
the number of shape functions calculated from Table \ref{tablica_1} 
\item
the number of Gauss points according to FEM accuracy considerations in Section \ref{section_accuracy}
\end{itemize}

The local memory (LS) available for each SPE, with the size of 256kB, is used for storing not only the data but also the code.
From Table \ref{tablica_3} it can be seen that several options have to be considered:
\begin{itemize}
\item
for low order approximations (up to 4) it should be possible to store precomputed values of shape functions and their derivatives at all integration points  in LS. The time for retrieving values from local memory is shorter than the time for recalculating the values at each integration point
\item
for higher orders, not only the values of shape functions and their derivatives at all integration points do not fit into LS, but also the resulting element stiffness matrix may be too large to be stored in one local array
\end{itemize}

Due to this fact, a modification to the original algorithm is introduced. The whole stiffness matrix is split into several parts and calculations are performed for subsequent parts. The loop over parts of stiffness matrix appears outside the loop over integration points. This means that calculations outside the double loop over shape functions (steps 6, 7, 8 in Algorithm \ref{alg_seq}), that are performed only once for each integration point in the original algorithm, are now repeated several times (as many times as there are parts of stiffness matrix). In that way, the small local memory of SPEs causes the increase in execution time for the numerical integration algorithm.

 \begin{table}[t!]
 \begin{center}
 \resizebox{13.5cm}{!}{
 \begin{tabular}{|c|l|r|r|r|r|r|r|r|}
 \hline
 Number & Computed &  \multicolumn{7}{|c|}{Degree of approximating polynomials p}  \\
 of operations & quantities & \textbf{1} & \textbf{2} & \textbf{3} & \textbf{4} & \textbf{5} &\textbf{6} & \textbf{7}   \\
 \hline
 at single &$\hat{\phi}_i$, $\frac{\partial \hat{\phi}_i}{\partial \bfx}$
 &	135&	378&	814&	1500&	2493&	3850&	5626\\
 \cline{2-9}
 integration& $\bfJ_{\bfT_e}$, $\bfJ_{\bfT_e^{-1}}$
 &	220&	220&	220&	220&	220&	220&	220\\
 \cline{2-9}
 point&$\bfA^e$
 &	252&	2268&	11200&	39375&	111132&	268912&	580608\\
 \hline
 for the whole& $\hat{\phi}_i$, $\frac{\partial \hat{\phi}_i}{\partial \bfx}$
 & 0.81 &	6.80 &	39.07 &	120.0 &	373.9 &	889.3 & 1891\\
 \cline{2-9}
algorithm&$\bfJ_{\bfT_e}$, $\bfJ_{\bfT_e^{-1}}$
 &1.32&	3.96 &	10.56&	17.60 &	33.00 &	 50.82 &	73.92 \\
 \cline{2-9}
 (in thousands)&$\bfA^e$
 & 1.51 &	40.82 &	537.6 &	3150 & 16669 & 62118 & 195084\\
 \hline
 \end{tabular}
 }
 \end{center}
 \caption{The number of operations for different steps in the numerical integration algorithm for 3D prismatic elements with different   degrees of approximating polynomials.}
 \label{tablica_2}
 \end{table}
  
For higher orders of approximation, as it was already discussed, two options exist for obtaining shape function values: they are either retrieved from global memory or calculated by SPEs. 
In order to better assess the costs associated with these alternatives, Table \ref{tablica_2} presents the numbers of operations for the numerical integration algorithm, at a single integration point and for all integration points,  for our example problem, according to the assumptions from section \ref{section_num_int}.

The number of operations for computing shape function values is calculated assuming that first, monomials in vertical direction and, separately, monomials in horizontal direction (following typical orientation of prismatic elements with triangular bases) are computed in 
$3*(p+1)*(p+2)/2+6*(p+1)$ operations. Then monomials for 3D shape functions and their derivatives are computed, according to tensor product rules, and multiplied by suitable coefficients. In our case, this procedure requires 19 operations per shape function. 

The number of operations for calculating Jacobian matrices is taken directly from the source code. The values associated with computing the entries of $\bfA^e$ are calculated assuming 7 operations per each stiffness matrix entry.

It can be seen that the number of operations per single shape function is less than 10 and hence slow transfers from global memory (with maximal bandwidth 25.6 GB/s for all SPEs) lead to worse overall execution times than local calculations. This result was confirmed also during computational tests and hence the option for retrieving shape functions from global memory is not discussed later in the paper.



 
 

\begin{figure}[t!]
\centering
\includegraphics[width=13.7cm]{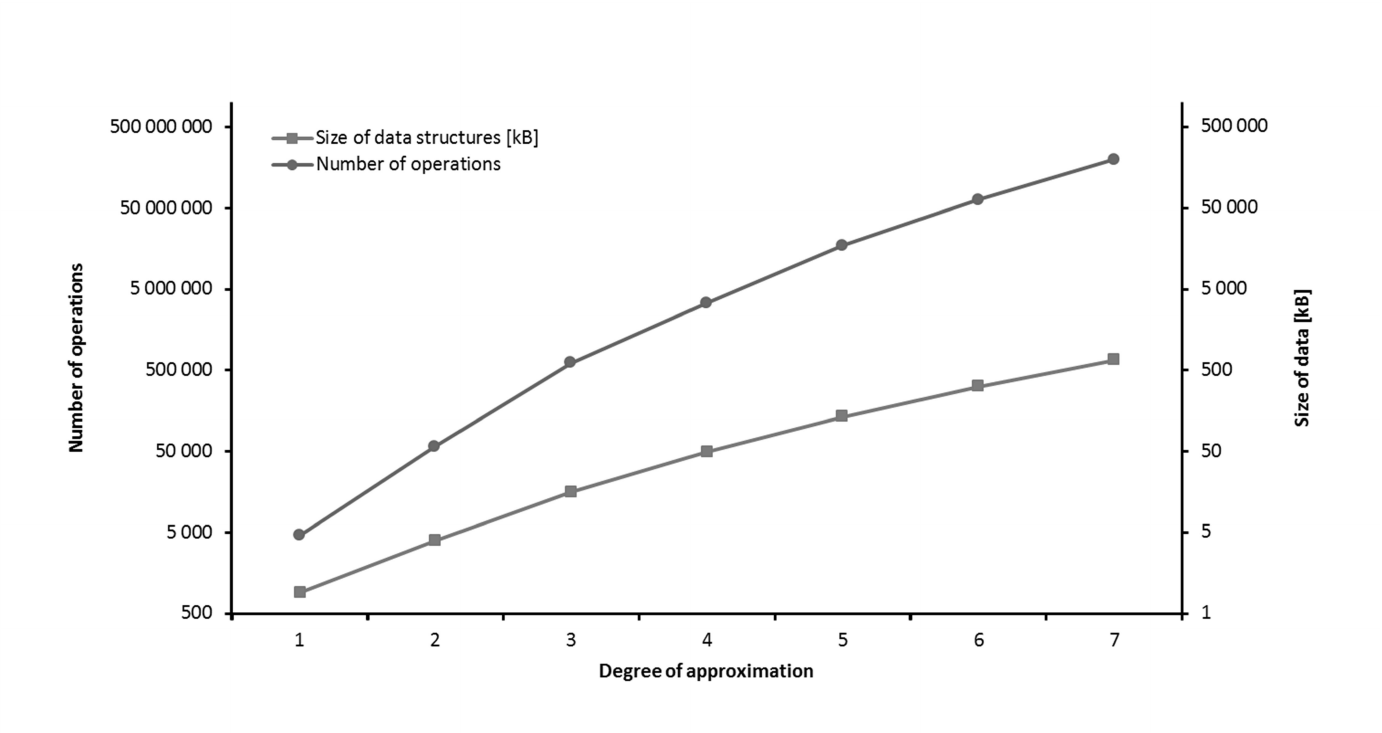}
\caption{The total number of operations for the numerical integration algorithm and the size of data structures used for 3D prismatic elements as a function of the order of approximation $p$.}
\label{obrazek2}
\end{figure}

Tables \ref{tablica_3} and \ref{tablica_2} are illustrated in Fig.~\ref{obrazek2} (observe logarithmic vertical scales). The faster growth of the number of operations as compared to the size of data structures, for increasing order of approximation, is clearly visible. From Table \ref{tablica_3} it can be seen that for higher orders of approximation the transfer from/to global memory is dominated by writing the computed element stiffness matrices. Since the number of operations for each computed entry grows for growing order $p$, one can expect the diminishing influence of slow speed of global memory transfers on the total execution time of the algorithm. Hence, the overall performance should depend more and more on the effective execution of the double loop over shape functions in Algorithm \ref{alg_seq}.

The performance in executing these loops depends on the proper mapping of computations to the last considered parallelization level, the level of SPE execution units. At this level two mechanisms are exploited: pipelined execution and SIMD capabilities of SPEs. The use of vector operations on 128-bit registers (corresponding to \textit{float4} vectors in OpenCL codes) is necessary in order to obtain high performance for PowerXCell processor.

\begin{algorithm}
\caption{Domain decomposition (DD) version of finite element numerical integration algorithm for PowerXCell processor.}
\label{alg_DD}
\begin{algorithmic}[1]
\STATE read quadrature data, $\bfxi$ and $w$, for all integration points, from global memory to local array
\STATE {\em (variant DD\_SR)} read values of shape functions and their derivatives with respect to local element coordinates, for all integration points, from global memory to local arrays
\FOR{$ielem=1$ \TO $nr\_elems\_in\_subset$}
\STATE read geometry data for element from global memory to local array
\FOR{$ipart=1$ \TO $nr\_parts\_of\_stiffness\_matrix$}
\STATE initialize part of element stiffness matrix $\bfA^e$ in local memory
\FOR{$I=1$ \TO $N_{\rm{I}}$} 
\STATE read from local arrays {\em (variant DD\_SR)} or compute {\em (variant DD\_SC)} values of shape functions and their derivatives with respect to local element coordinates
\STATE calculate Jacobian matrix, its determinant ($\det \bfJ_{\bfT_e}(\bfxi^I)$) and its inverse
\STATE calculate derivatives of element shape functions with respect to physical coordinates $\bfx$, $\frac{\partial \hat {\phi}_j}{\partial x_k}(\bfxi^I)$ and store in three arrays of \textit{float4} vectors (one array for each derivative, one vector for four consecutive shape functions)
\FOR{$i=1$ \TO $N_{\mathrm{sh}}$}
\STATE store derivatives $\frac{\partial \hat {\phi}_i}{\partial x_k}(\bfxi^I)$ in registers
\FOR{$j=1$ \TO $N_{\mathrm{sh}}$/4 }
\STATE update subsequent \textit{float4} vector (associated with index $j$) storing entries of $\bfA^e$, using registers and derivatives of shape functions from local array of \textit{float4} vectors, according to suitably modified formula (\ref{wzor2})
\ENDFOR
\ENDFOR
\ENDFOR
\STATE write part of element stiffness matrix to global memory
\ENDFOR
\ENDFOR
\end{algorithmic}
\end{algorithm}

In order to utilize SIMD execution units \textit{float4} vector variables have to be employed. The first choice is to apply the data decomposition approach, with several subsequent entries of the stiffness matrix forming a single vector. In this case Algorithm \ref{alg_seq} is transformed to its first parallel version presented as Algorithm \ref{alg_DD} (and denoted by DD - {\em data decomposition} version). There are two variants -- one with shape functions read from local arrays (DD\_SR, only for orders of approximation up to 4) and the second with shape functions computed locally by SPEs (DD\_SC). The algorithm is executed by each SPE thread, for a subset of elements assigned to it.

\begin{algorithm}
\caption{Finite element numerical integration algorithm for PowerXCell processor with vectorization based on weak formulation (WF version, only operations different than in the DD version are indicated).}
\label{alg_WF}
\begin{algorithmic}[1]
\STATE ...
\FOR{$ielem=1$ \TO $nr\_elems\_in\_subset$}
\STATE ...
\FOR{$ipart=1$ \TO $nr\_parts\_of\_stiffness\_matrix$}
\STATE ...
\FOR{$I=1$ \TO $N_{\rm{I}}$} 
\STATE ...
\STATE calculate derivatives of element shape functions with respect to physical coordinates $\bfx$, $\frac{\partial \hat {\phi}_j}{\partial x_k}(\bfxi^I)$ and store in array of \textit{float4} vectors (one vector for three derivatives of single shape function)
\FOR{$i=1$ \TO $N_{\mathrm{sh}}$}
\STATE store derivatives $\frac{\partial \hat {\phi}_i}{\partial x_k}(\bfxi^I)$ in vector register (\textit{float4} variable)
\FOR{$j=1$ \TO $N_{\mathrm{sh}}$ }
\STATE store derivatives $\frac{\partial \hat {\phi}_j}{\partial x_k}(\bfxi^I)$ in vector register (\textit{float4} variable)
\STATE update vector register storing three parts of $A^e_{ij}$ related to different space dimensions using previously prepared vector registers, according to suitably modified formula (\ref{wzor2})
\ENDFOR
\ENDFOR
\ENDFOR
\STATE sum up three components for each entry $A^e_{ij}$ and write part of element stiffness matrix to global memory
\ENDFOR
\ENDFOR
\end{algorithmic}
\end{algorithm}

We consider also another form of employing vector operations when executing a double loop over shape functions. In Algorithm \ref{alg_DD} for each stiffness matrix entry the calculations in a loop over space dimensions, steps 11-13 in Algorithm \ref{alg_seq}, are executed sequentially. In the second variant of numerical integration algorithm for PowerXCell processor we perform these operations in parallel. Instead of summing up the contributions from different space dimensions for each integration point, we calculate them separately, in different components of vectors forming a suitably created array. These components are summed up just before writing the part of stiffness matrix to global memory. This version of algorithm is denoted by WF (parallelization based on {\em weak formulation}) and also exists in two variants, WF\_SR and WF\_SC. The disadvantage of the WF version is the increase of necessary storage, since data for each direction are stored independently during calculations. Algorithm \ref{alg_WF} presents the WF version of numerical integration for PowerXCell, with parts identical as in the DD version omitted.


The final step of mapping the algorithm to hardware resources consists in proper utilization of execution pipelines. Since there are no explicit mechanisms for doing that, we achieve this goal by performing software optimizations. The results of optimizations are than checked using 
 the {\em spu\_timing} tool provided with the CBE SDK for assembler code analysis. The tool presents operations performed in each clock cycle by each of SPE pipelines. The inspection of the gathered data indicates how far is the algorithm from reaching the theoretical peak performance of the hardware.
 
The main optimization technique that we employ is loop unrolling, applied to both loops over shape functions. The WF version of algorithm allows for unrolling with larger factor than the DD version (at least for lower orders of approximation, where the number of entries of stiffness matrix is small). For the DD version we unroll both loops with the factor 2, while for the WF algorithm we unroll the external loop with the factor 2 and the internal loop with the factor 8. 

\subsection{Implementation for OpenCL platform}




The OpenCL programming model requires the creation of a special PPE host code that manages the execution of kernels on SPEs.
The associated flow of control begins with PPE creating OpenCL context and loading the source code for SPE cores. 
Then PPE computes characteristics of a particular instance of numerical integration algorithm. This characteristics include sizes of data structures for input data, such as 
nodal coordinates and integration quadratures. It also compares the available local SPE resources with the space necessary to hold the element stiffness matrix. It takes into account the workspace for all shape functions and their derivatives, as well as the fact that stiffness matrices are properly padded.
If the local element matrix is too large, PPE computes parameters of the matrix division into horizontal groups of rows.  The main computed characteristics of the execution for our model problem and different orders of approximation are gathered in Table~\ref{CBE_execution_characteristics}. 

The data in the table corresponds to the WF algorithm, with the last row containing the numbers of actually performed operations for the WF\_SC version of the algorithm. These numbers exceed the numbers of operations from Table~\ref{tablica_2}. There are two reasons -- one is the padding of stiffness matrices and the second is the fact of performing vector operations on 128-bit vectors with only three 32-bit values packed. Hence for each 3 operations performed by scalar cores, like x86 cores, a SPE is performing 4 equivalent operations.

Similarly, each SPE is transferring the content of its 128-bit registers to/from the local store, with only 96 bits holding data relevant to calculations. Thus the effective performance of the processor is lower than the performance of actually performed operations.

The calculations of execution characteristics for the DD version are performed in a similar way and give similar results. The main difference appears in the last row of data, where the number of operations for the DD algorithm approaches the numbers in Table \ref{tablica_2}.

\begin{table}
\begin{center}
{\footnotesize
\begin{tabular}{|c|c|c|c|c|c|c|c|}
\hline
\hline
& p=1 & p=2 & p=3 & p=4 & p=5 & p=6 & p=7 \\
\hline
The size of padded &&&&&&& \\
stiffness matrix [kB] 
& 0.19 & 1.73 & 6.40 & 24.0 & 64.5 & 157 & 332 \\
\hline
The number of parts&&&&&&& \\
of the stiffness matrix
& 1 & 1 & 1 & 1 & 2 & 5 & 9 \\
\hline
The number of rows in the &&&&&&& \\
part of the stiffness matrix
& 8 & 24 & 40 & 80 & 64 & 40 & 32 \\
\hline
The size of the part &&&&&&& \\
of the stiffness matrix [kB]
& 0.77 & 6.91 & 25.60 & 97.28 & 129.0 & 125.4 & 147.4 \\
\hline
The size of workspace  &&&&&&& \\
for shape functions [kB]
& 0.18 & 0.43 & 0.85 & 1.50 & 2.41 & 3.64 & 5.23 \\
\hline
The actual number  &&&&&&& \\
 of operations ($\times 10^{-6}$)
& 0.004 & 0.072 & 0.657 & 3.965 & 20.00 & 76.42 & 238.0 \\
\hline
\hline
\end{tabular}
}
\end{center}
\caption{The characteristics of PowerXCell calculations for numerical integration of element stiffness matrices using the WF algorithm for the model problem and different orders of approximation $p$.}
\label{CBE_execution_characteristics}
\end{table}

After computing execution characteristics, PPE prepares buffers for input data such as coordinates of element nodes, weights and coordinates of quadrature points, as well as some other parameters of execution. Using special OpenCL functions the buffers are made available to kernels running on SPEs. PPE prepares also buffers on SPE side, for shape functions workspace and for the part of the stiffness matrix. Next the space for the element stiffness matrix being the output of computations is created. 
Finally, OpenCL procedure for running the kernel is called and SPEs start their calculations. OpenCL runtime environment ensures the proper passing of arguments being the addresses of input, output and workspace buffers.

As a result of computations element stiffness matrices are created and stored in global memory. This is achieved by associating OpenCL global memory with PowerXCell DRAM. Hence, explicit OpenCL transfers are not necessary and the control of memory operations is determined by the OpenCL compiler. Nevertheless, the writes to global memory are performed in large chunks, using the values stored in local memory for the parts of stiffness matrices. In that way, we try to allow the compiler to optimize memory transfers. The computed element stiffness matrices reside in host memory, ready to be assembled by different software components of a finite element code.

\section{Numerical experiments}
To test our algorithms, we used one QS22 computational node of IBM BladeCenter QS22/LS21 server equipped with two IBM PowerXCell 8i processors with 3.2 GHz frequency and 8GB RAM.
 We employed IBM OpenCL SDK 0.3 with gcc as the compiler.
 We always used one PPE and 16 SPEs belonging to both server processors.

For comparison, a CPU optimized numerical integration algorithm was also run on a laptop computer with a modern Intel Core i5--2520M processor (2.5 GHz). The code was compiled by the  Intel C compiler (icc) with the -O3 optimization flag, that switches on code vectorization. 

The task performed was the creation of element matrices that occupy approximately 500 MB of global DRAM memory. This corresponds to a typical scenario in large scale distributed calculations where the number of elements for a given computational node is memory limited (the particular value for the limit is usually determined by linear solver requirements). We assumed the constant size for all stiffness matrices which, considering the fact that for growing orders of approximation the size of element stiffness matrices grows (see Table \ref{tablica_2}), results in the decreasing number of elements per kernel (and per one PowerXCell core) for increasing $p$. One of the consequences is the fact that the overhead of calculations (which is almost the same for each $p$ and a single kernel invocation) amortizes over less elements for growing $p$.

\begin{table}
 \begin{center}
 {\footnotesize
  \begin{tabular}{|r|r|r|r|r|r|r|r|}
 \hline
 \hline
 Execution time & p=1 & p=2 & p=3 & p=4 & p=5 & p=6 & p=7 \\
 \hline
 \hline
 {Single Sandy Bridge core}
 & 0.809 & 6.417 & 43.00 & 242 & 1070 & 3659 & 11505\\
 \hline
 \hline
 {PowerXCell, 16 SPEs} \\
\hline
 { - initialization overhead}
 & 0.123 & 0.741 & 2.60 & 9.69 & 25.7 & 62 & 132\\
\hline
 {DD\_SC -- total execution}
& 0.697&	4.841&	21.65&	75.00&	228.9&	773&	2691 \\
 \hline
 {DD\_SR -- total execution}
&0.546&	3.271&	17.10&	63.59& -- & -- & -- \\
 \hline
 {WF\_SC -- total execution} 
 &0.601 &	3.944 &	17.63 &	68.72 &	276.3 &	1064 &	3348 \\
\hline
 {WF\_SR -- total execution}
 &0.458	 &2.755	 &13.95 &	63.32& -- & -- & -- \\
\hline
 \hline
 {PowerXCell, 1 SPE} \\
\hline
 { - initialization overhead}
& 0.119	&0.703&	2.51&	9.34&	24.9&	59&	127 \\
 \hline
 {DD\_SC -- total execution} 
&8.685	&61.83	&291.2	&992	&3115	&11077	&40584 \\
 \hline
 {DD\_SR -- total execution} 
& 6.265	&47.16	&217.8	&815.1 & -- & -- & -- \\
 \hline
 {WF\_SC -- total execution} 
 &7.256	 &48.04 &	229.4 &	900.1	 &3893 &	15738 &	50987 \\
 \hline
 {WF\_SR -- total execution} 
 &4.989	 &29.09 &	166.6	 &792.9& -- & -- & -- \\
  \hline
 \hline
 \end{tabular}
 }
 \end{center}
 \caption{Numerical integration execution times (in microseconds) for the model problem, a single element  stiffness matrix, different integration algorithms and different orders of approximation.}
 \label{times}
 \end{table}

The results of experiments for the model problem and different orders of approximation are presented in Tables \ref{times} and \ref{gigaflops}. 
Table \ref{times} shows the execution times for a single element, calculated by dividing the execution times for all elements by the number of elements.
The table presents raw execution times measured by system tools (for OpenCL measured at host side). The results for the PowerXCell processor are split into different variants of implementation and additionally into the initialization phase (that takes a significant portion of the execution time for $p=1$) and the total execution that includes the OpenCL initialization and the kernel operations (calculations and global memory transfers).

\begin{figure}[t!]
\centering
\includegraphics[width=13cm]{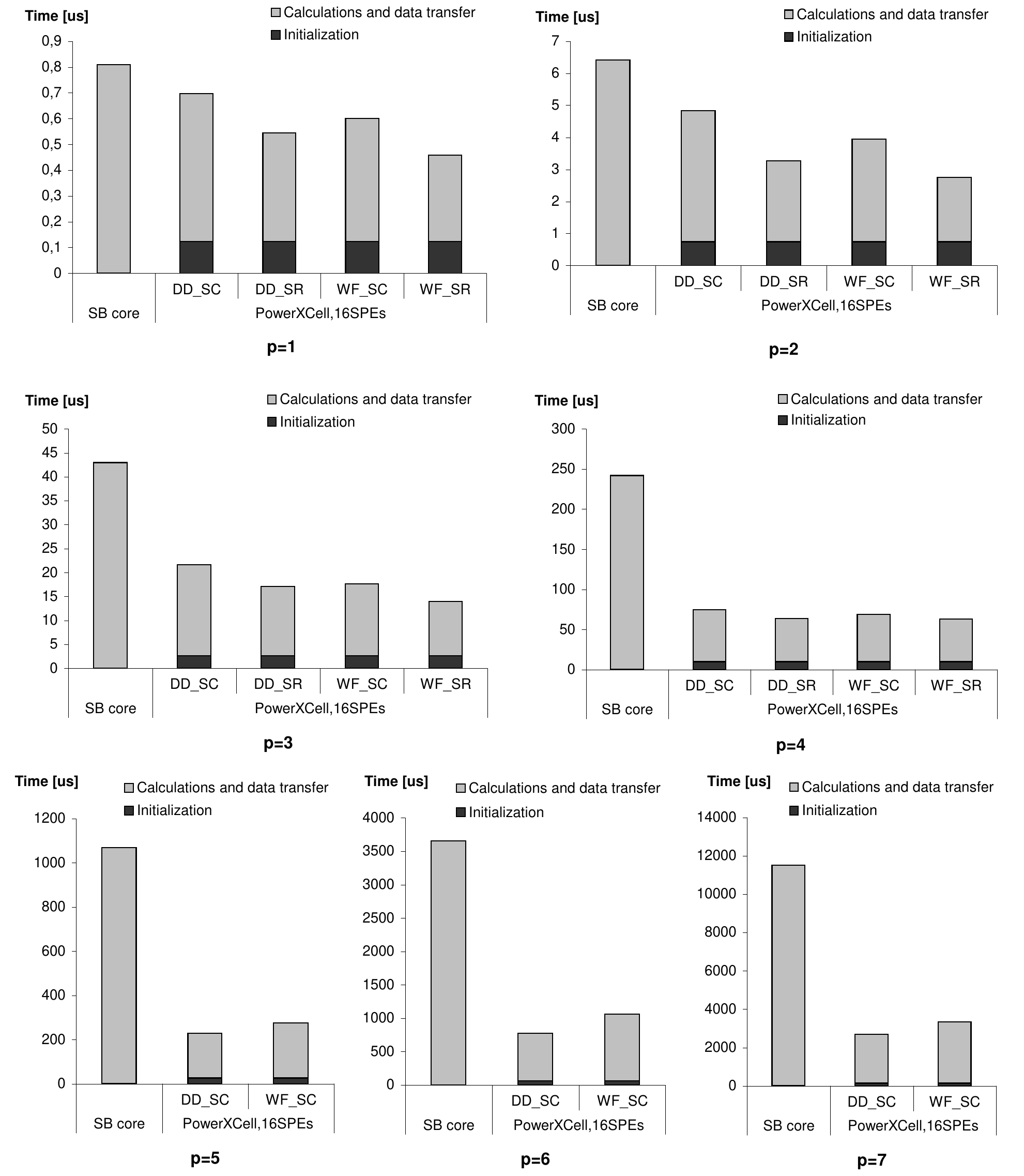}
\caption{Numerical integration execution times (in microseconds) for the model problem, a single element  stiffness matrix, different integration algorithms and different orders of approximation. Times for PowerXCell execution are split into OpenCL initialization and kernel execution, the latter including data transfers.}
\label{fig_times}
\end{figure}

The results are illustrated in Fig.~\ref{fig_times}. It can be seen that, for low order approximations, PowerXCell does not offer significant improvements, even with respect to a single core of modern processors. With the increasing order $p$ and the growing ratio of calculations to global memory transfers the advantages of PowerXCell become more evident. 

 \begin{table}
 \begin{center}
 {\footnotesize
 \begin{tabular}{|r|r|r|r|r|r|r|r|}
 \hline
 \hline
 Performance in GFlops & p=1 & p=2 & p=3 & p=4 & p=5 & p=6 & p=7 \\
 \hline
 \hline
 \parbox [t]{4.0cm}{Single Sandy Bridge core}
 & 4.50 & 8.04 & 13.66 & 13.58 & 15.96 & 17.23 & 17.13 \\
 \hline
 \parbox [t]{4.0cm}{Example 4-core processor }
 & 18.02 & 32.16  & 54.63 & 54.34 & 63.84 & 68.94 & 68.51 \\
  \hline
  \hline
 \parbox [t]{4.0cm}{PowerXCell, 16 SPEs } \\
 \hline
 \parbox [t]{4.0cm}{DD\_SC -- internal }
  &8.16	 &17.33	 &33.40	 &54.38	 &86.77 &	93.26 &	79.86 \\
\hline
 \parbox [t]{4.0cm}{DD\_SC -- external }
 &5.22	 &10.66	 &27.12	 &43.83	 &74.61	 &81.51	 &73.22 \\
 \hline
 \hline
 \parbox [t]{4.0cm}{DD\_SR -- internal }
 & 9.66&	21.39	&43.09	&65.34& -- & -- & -- \\
 \hline
 \parbox [t]{4.0cm}{DD\_SR -- external }
& 6.67	&15.77	&34.34	&51.69& -- & -- & -- \\
 \hline
 \hline
 \parbox [t]{4.0cm}{WF\_SC -- internal }
  & 11.43 & 32.50 & 64.29 & 100.39 & 118.46 & 112.32 & 108.53 \\
 \hline
 \parbox [t]{4.0cm}{WF\_SC -- external }
& 6.06 & 13.08 & 33.29 & 47.84 & 61.80 & 59.22 & 58.84 \\
 \hline
 \hline
 \parbox [t]{4.0cm}{WF\_SR -- internal }
  &14.43	 &40.14 &	84.29	 &111.52& -- & -- & -- \\
 \hline
 \parbox [t]{4.0cm}{WF\_SR -- external }
 &7.94	&18.72	&42.07	&51.92& -- & -- & -- \\
 \hline
 \hline
 \end{tabular}
 }
 \end{center}
 \caption{Performance achieved by x86 and PowerXCell processors for the model problem, a single element  stiffness matrix, different integration algorithms and different orders of approximation (explanation in text).}
 \label{gigaflops}
 \end{table}

The second table presents characteristics computed on the basis of measured quantities. The performance of the single Sandy Bridge core is calculated and then extrapolated  for a typical contemporary quad-core processor, assuming perfect scalability with the growing number of cores. For each algorithm run on the PowerXCell processor two results are presented. The first is the performance as observed by an external user -- the number of operations theoretically necessary to obtain results (taken from Table~\ref{tablica_2}, the same as for the Sandy Bridge processor) is divided by the total time of OpenCL execution, measured on host side and including initialization phase. The second value, denoted ''internal'', is computed by dividing the number of operations actually performed by the hardware by the time of pure kernel execution. The difference between the two values for the WF algorithms is associated not only with the existence of initialization phase, but also with the fact that the processor performs more operations than it is required by the original, sequential algorithm.

\begin{figure}[t!]
\centering
\includegraphics[width=13cm]{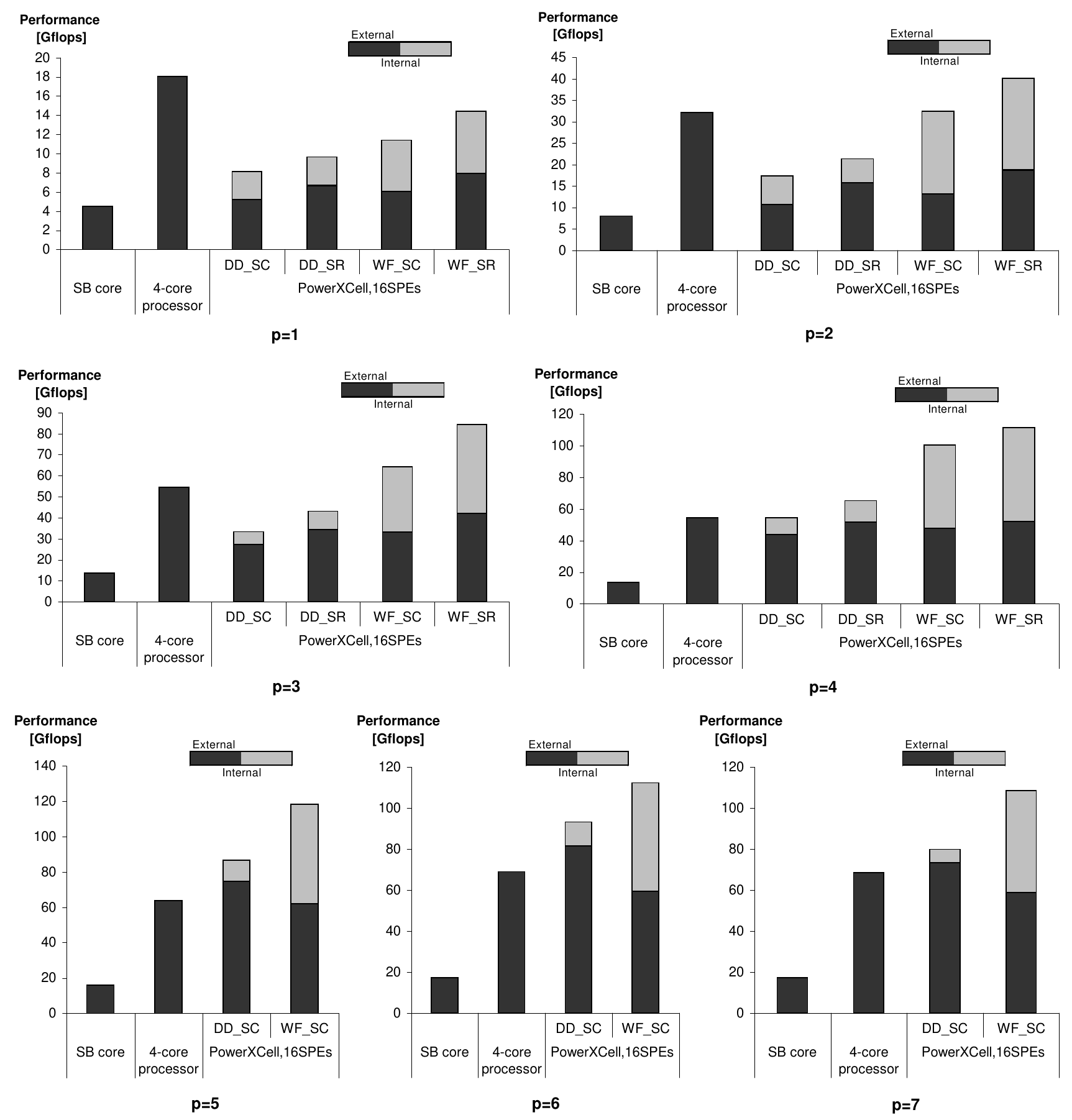}
\caption{Performance characteristics for the model problem, a single element  stiffness matrix, different integration algorithms and different orders of approximation. Two kinds of performance are indicated for PowerXCell processor: \textit{internal} -- reflecting the actual performance of hardware executing instructions and \textit{external} -- computed as the number of operations required by the standard sequential algorithm, divided by the PowerXCell execution time. }
\label{fig_perf}
\end{figure}

Figure \ref{fig_perf} shows graphically the performance results. It can be seen that, despite the fact that Intel architecture represents much more recent design, the performance of both platforms is comparable. For the Sandy Bridge core, the high performance was achieved by exploiting vector operations and single precision calculations. In the domain of general purpose scientific computing the ranges of performance obtained for this core and higher orders approximations should be considered very high.

For the PowerXCell processor the results can be considered satisfactory for higher orders of approximation. 
Similarly to x86 Sandy Bridge cores, SPE cores made good use of the increasing ratio of calculations performed inside vectorized loops to the calculations performed outside the loops. 
Despite substantial global memory transfers and the existence of parts of algorithms where neither FMA operations nor vectorization cannot be applied (and hence the performance drops at least by the factor of 8), the performance of pure calculations reaches more than 79 GFlops (19\% of theoretical maximum) for the DD\_SC version and more than 108 GFlops (26\% of theoretical maximum) for the WF\_SC variant.

For lower orders of approximation, the overhead associated with the initialization of calculations and the transfers from/to global DRAM memory significantly slows down execution.
The number of operations for a single transferred entry of stiffness matrices is too small, considering also the fact that scalar operations, outside the double loop over shape functions, form high percentage of all calculations.


\begin{figure}[t!]
\centering
\includegraphics[width=13cm]{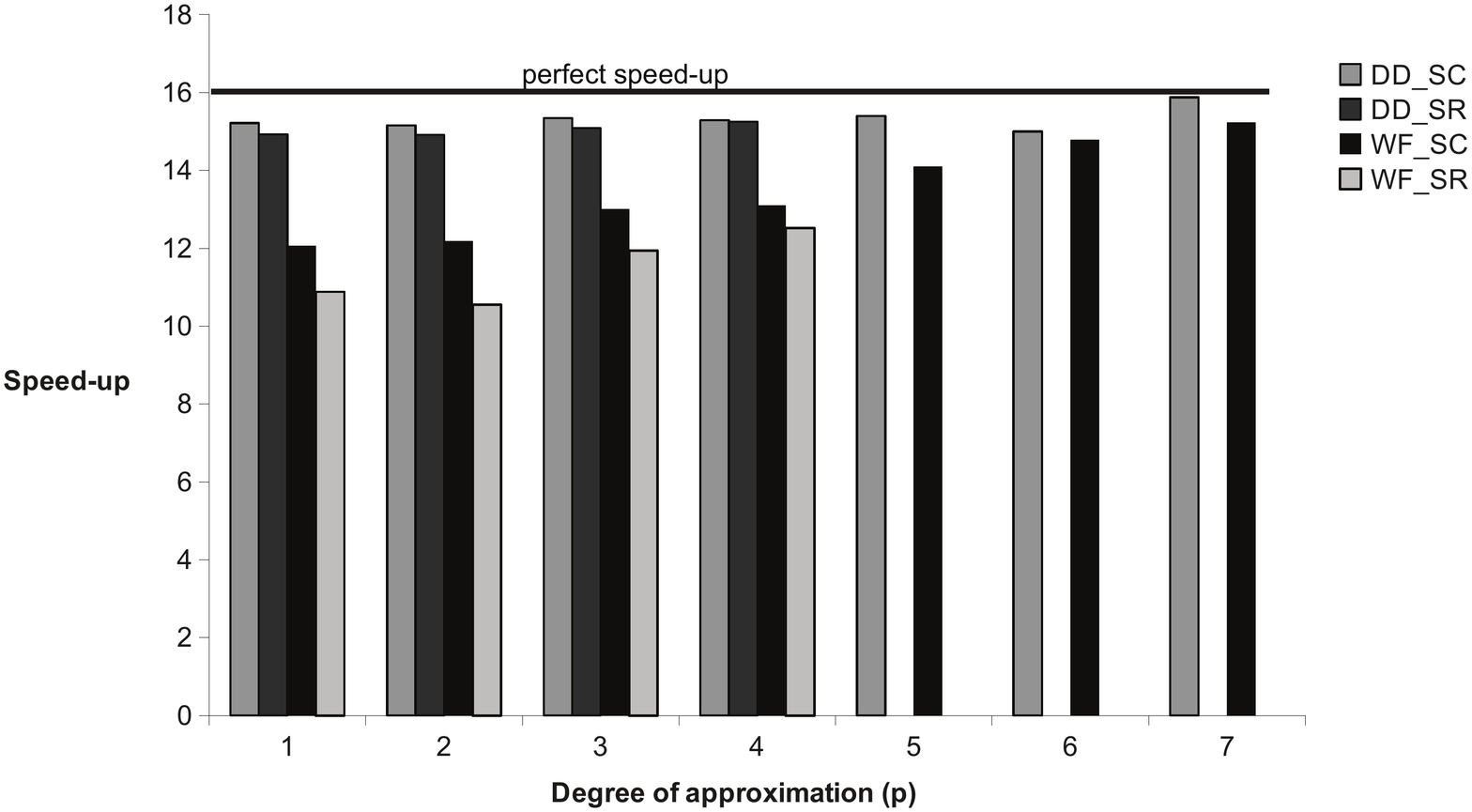}
\caption{Speed-up achieved by 16 SPEs of the PowerXCell processors for the WF version of numerical integration algorithm for the model problem and different orders of approximation. }
\label{fig_speed_up}
\end{figure}

One more characteristic of the parallel execution of numerical integration on PowerXCell processor,  that can be deduced form data in Table \ref{times}, is the classical measure of speed-up. It is depicted in Fig.~\ref{fig_speed_up}. It can be seen that for 16 cores, the speed-up ranges from approx.~11 to almost 16 (with parallel efficiency always above 68\% and for $p=7$ reaching more than 95\%).

Comparing the different versions of the implementation of numerical integration algorithm, one can conclude that the SR variants (with shape functions read from global memory) give always better performance results, while the SC versions (with shape functions computed locally) show better flexibility, by requiring much less memory resources. 

The choice between the DD and WF variants is definitely problem dependent. We have developed the WF algorithms to present the possibility of efficient parallelization that does not use data decomposition approach associated with element stiffness matrices. For certain situations (e.g. when data decomposition is used for spreading calculation over different cores, as is the case for GPUs), the option of parallelization based on a particular form of weak statement may become optimal. For our example case of Laplacian term, both approaches turned out to be roughly equivalent.

\section{Conclusions}
The performance results presented for the finite element numerical integration algorithm running on the PowerXCell processor prove that the algorithm can be successfully ported to multi-core processors with manually managed memory hierarchy and vector execution units. The obtained range of performance numbers shows that in many situations high utilization of vector capabilities can be achieved. This seems to be an important conclusion in the light of a recently observed trend to equip standard processor cores with wide vector registers and execution units.

Another conclusion is that OpenCL can be used for relatively simple porting of scientific codes to complex heterogeneous multi-core architectures, such as CBE. Moreover, 
OpenCL allows one to obtain high performance code, thanks to the support of explicit memory hierarchy management and vector operations. Again, the experience gathered with PowerXCell can be used in designing code for other types of processors, sharing architectural details with PowerXCell, such as e.g. Xeon Phi. 

As a further step of our investigations we plan to combine the experiences gathered in porting numerical integration procedures to PowerXCell processor and GPUs \cite{ppam_09_gpu,imcsit_10,iccs_10_gpu,cmwa_12_ni_GPU} and try to design parametrized kernels for different architectures, including new Xeon Phi co-processors.

In a more general context, we see our efforts in transferring the numerical integration algorithm to the PowerXCell processor as a step towards better understanding the relation between the software, in our case the finite element numerical integration algorithm, and the hardware of modern multi-core processors. We hope to utilize this knowledge in porting the whole finite element code to other modern multi- and many-core processors. 

\subsubsection*{Acknowledgments.}
 The support of this work by the Polish National Science Center under grant no DEC-2011/01/B/ST6/00674 is gratefully acknowledged.











\end{document}